\documentclass[12pt,preprint]{aastex}

\def\beq{\begin{eqnarray}}
\def\eeq{\end{eqnarray}}
\def\d{{\rm d}}
\def\a{\alpha}
\def\Rg{R_{\rm g}}
\def\cs{c_{\rm s}}
\def\Om{\Omega}
\def\OmK{\Omega_{\rm K}}
\def\dM{\dot M}
\def\dm{\dot m}
\def\Lv{L_{\nu}}
\def\Lvv{L_{\nu\bar{\nu}}}

\def\Qvis{Q_{\rm vis}}
\def\Qadv{Q_{\rm adv}}
\def\Qv{Q_{\nu}}

\def\MBH{M_{\rm BH}}

\def\ergs{ergs~s$^{-1}$}

\shorttitle{NDAF Models for GRBs}
\shortauthors{Gu, Liu, \& Lu}

\begin{document}

\title{Neutrino-Dominated Accretion Models for Gamma-Ray Bursts:
Effects of General Relativity and Neutrino Opacity}

\author{Wei-Min Gu, Tong Liu, and Ju-Fu Lu}

\affil{Department of Physics
and Institute of Theoretical Physics and Astrophysics, \\
Xiamen University, Xiamen, Fujian 361005, China}

\email{guwm@xmu.edu.cn}

\begin{abstract}
We first refine the fixed concept in the literature that the usage of
the Newtonian potential in studies of black hole accretion is invalid
and the general relativistic effect must be considered.
As our main results, we then show that the energy released by neutrino
annihilation in neutrino-dominated accretion flows is sufficient
for gamma-ray bursts when the contribution from the optically thick
region of the flow is included, and that in the optically thick region
advection does not necessarily dominate over neutrino cooling because
the advection factor is relevant to the geometrical depth rather than
the optical depth of the flow.
\end{abstract}

\keywords{accretion, accretion disks --- black hole physics ---
gamma rays: bursts --- neutrinos}

\section{Introduction}

The fireball shock model
(see, e.g., M\'esz\'aros 2002 and Zhang \& M\'esz\'aros 2004 for reviews)
has been widely accepted to interpret the gamma-ray
and afterglow emitting of gamma-ray bursts (GRBs).
Despite the successes of this phenomenological model, the central engine
of the relativistic fireball is not yet well understood.
Most popular models for the energy source of GRBs are in common
invoking a hyperaccreting black hole. Accretion models in this context
were first considered by Narayan, Paczy\'nski, \& Piran (1992),
and have been recently discussed by Popham, Woosley, \& Fryer (1999,
hereafter PWF), Narayan, Piran, \& Kumar (2001), Kohri \& Mineshige (2002),
Di Matteo, Piran, \& Narayan (2002, hereafter DPN), and
Kohri, Narayan, \& Piran (2005).

PWF introduced the concept of neutrino-dominated accretion flows (NDAFs)
and showed that the energy released by neutrino annihilation was
adequate for GRBs. Their calculations, however,
were based on the assumption that the flow is optically thin for
neutrinos. As pointed out by themselves, this assumption breaks down
for the mass accretion rate $\dM \ga 10 \ M_\sun$ s$^{-1}$.
They mentioned that their estimate of the neutrino annihilation
luminosity $\sim 2 \times 10^{53}$ \ergs
(see their Table~3) for $\dM = 10 \ M_\sun$ s$^{-1}$ 
should be taken as an upper limit,
and the actual luminosity could be as much as a factor of 5 lower,
i.e., $\sim 4 \times 10^{52}$ \ergs.
The NDAF model was reinvestigated by DPN, in which a bridging formula was
adopted for calculating neutrino radiation in
both the optically thin and thick cases.
They showed that for $\dM > 0.1 \ M_\sun$ s$^{-1}$
there exists an optically thick inner region in the flow;
and argued that for $\dM \ga 1 \ M_\sun$ s$^{-1}$
neutrinos are sufficiently trapped and energy advection
becomes the dominant cooling mechanism,
resulting in the maximum luminosity of neutrino annihilation which is only
$\sim 10^{50}$ \ergs (see their Fig.~6).
Thus they claimed that the NDAF model cannot account for GRBs.

How to understand the inconsistent results of PWF and DPN?
We note that PWF worked in the relativistic Kerr geometry,
but with the a priori assumption that neutrinos are optically thin;
whereas DPN calculated the optical depth for neutrinos,
but went back into the Newtonian potential and omitted totally
the neutrino radiation from the optically thick region.
The purpose of this {\sl Letter} is to try to update partly the
NDAF model. It is surely correct that the general relativistic
effect must be considered in studies of black hole accretion,
then we wish to know how important the effect of the neutrino opacity
is in determining the luminosity of an NDAF.

\section{Assumptions and equations}

For simplicity, a steady state axisymmetric black hole accretion flow
is considered as in PWF and DPN.
We adopt that the general relativistic effect of the central black hole
is simulated by the well-known pseudo-Newtonian potential introduced by
Paczy\'nski \& Wiita (1980, hereafter PW potential),
i.e., $\Phi = -G\MBH /(R-\Rg)$, where $\MBH$ is the black hole mass,
$R$ is the radius, and $\Rg = 2 G\MBH /c^2$ is the Schwarzschild radius.
Other assumptions about the flow are usual in the literature:
the angular velocity is approximately Keplerian, i.e.,
$\Om = \OmK = (GM_{\rm BH}/R)^{\frac{1}{2}}/(R-\Rg)$;
the vertical scale height of the flow is $H = \cs/\OmK$, where
$\cs = (P/\rho)^{\frac{1}{2}}$ is the isothermal sound speed,
with $P$ and $\rho$ being the pressure and mass density, respectively;
and the kinematic viscosity coefficient is expressed as $\nu = \a \cs H$,
where $\a$ is the constant viscosity parameter.

The basic equations describing the flow
consist of the continuity, azimuthal momentum,
and energy equations plus the equation of state.
The continuity equation is
\beq
\dM = -4 \pi \rho H R v \ ,
\eeq
where $v$ is the radial velocity.
With the assumption $\Om = \OmK$, the azimuthal momentum equation
is reduced to an algebraic form:
\beq
v = -\a \cs \frac{H}{R} f^{-1} g \ ,
\eeq
where $f = 1-j/(\Om R^2)$ and $g = -\d \ln \OmK/\d \ln R$, with the
integration constant $j$ representing the specific angular 
momentum (per unit mass) accreted by the black hole.
The equation of state is written as
\beq
P = P_{\rm gas} + P_{\rm rad} + P_{\rm deg} + P_{\nu} \ ,
\eeq
where $P_{\rm gas}$, $P_{\rm rad}$, $P_{\rm deg}$, and $P_{\nu}$
are the gas pressure from nucleons, radiation pressure of photons,
degeneracy pressure of electrons,
and radiation pressure of neutrinos, respectively.
The energy equation is written as
\beq
\Qvis = \Qadv + Q_{\rm photo} + \Qv \ .
\eeq
The viscous heating $\Qvis$ and the advective cooling $\Qadv$
(for a half disk above or below the equator) are expressed as
\beq
\Qvis = \frac{1}{4\pi} \dot M \Om^2 f g \ ,
\eeq
\beq
\Qadv = \rho H v T \frac{\d s}{\d R} \backsimeq - \xi v \frac{H}{R}T
( \frac{11}{3}aT^3 + \frac{3}{2} \frac{\rho k}{m_{\rm p}}
\frac{1+3X_{\rm nuc}}{4} + \frac{4}{3}\frac{u_{\nu}}{T} ) \ ,
\eeq
where $T$ is the temperature, $s$ is the specific entropy,
$X_{\rm nuc}$ is the mass fraction of free nucleons,
$u_{\nu}$ is the neutrino energy density,
and $\xi \propto -\d \ln s/\d \ln R$
is assumed to be equal to 1 as in DPN. The quantity
$Q_{\rm photo}$ is the cooling of the photodisintegration process,
and $\Qv$ is the cooling of the neutrino radiation. We adopt a
bridging formula for calculating $\Qv$, which is
valid in both the optically thin and thick cases.
Detailed expressions for $P_{\rm gas}$, $P_{\rm rad}$,
$P_{\rm deg}$, $P_{\nu}$, $Q_{\rm photo}$, $X_{\rm nuc}$,
$u_{\nu}$, and the bridging formula for $\Qv$ are given
in DPN.

Equations~(1-4) contain four independent unknowns
$\rho$, $T$, $H$, and $v$ as functions of $R$,
which can be numerically solved
with given constant parameters $\MBH$, $\dM$, $\a$, and $j$,
then all the other quantities can be obtained.
In the following calculations we fix
$\MBH = 3 M_\sun$ and $\a = 0.1$.

\section{Invalidity of the usage of the Newtonian potential}

Most previous calculations for NDAFs (e.g., Narayan et al. 2001;
DPN; Kohri et al. 2005) adopted the Newtonian potential and
did not take the integration constant $j$ into consideration.
Kohri \& Mineshige (2002) also used the Newtonian potential but
considered $j$. Only PWF worked in the relativistic Kerr geometry
as we mentioned already.
In this section we refine the fixed concept in the literature, i.e.,
the invalidity of the usage of the Newtonian potential.
We concentrate on three solutions corresponding to
PW potential with $j = 1.8 c\Rg$ (just a little less than
the Keplerian angular momentum at the last stable orbit,
$l_{\rm K}|_{3 \Rg} = 1.837 c\Rg$), the Newtonian potential
with $j = 1.2 c\Rg$ ($l_{\rm K}|_{3 \Rg} = 1.225 c\Rg$,
cf. Kohri \& Mineshige 2002), and the Newtonian potential with $j=0$
(DPN; Kohri et al. 2005), respectively.

The variation of the optical depth $\tau$ for neutrinos with $R$
is drawn in Figure~1(a), for which the dimensionless mass
accretion rate is $\dm \equiv \dM/(M_\sun$ s$^{-1}) = 1$.
The figure shows that the values of $\tau$ in the Newtonian potential
(the dotted and dashed lines) are significantly larger than
those in PW potential (the solid line).
The accretion flow in PW potential is completely optically thin
($\tau < 2/3$),
whereas for the Newtonian potential with $j=0$ there exists a wide
optically thick ($\tau > 2/3$) region of $R \la 15.4 \Rg$.
We believe that the results with PW potential are more
convincible since this potential is known to be a better description
for a nonrotating black hole than the Newtonian potential.
Our arguement can be further confirmed by Figure~1(b),
which shows the variation of $\eta_{\nu}$
with $\dm$, where $\eta_{\nu} \equiv \Lv/\dM c^2$ is the
efficiency of energy release by neutrino radiation (before annihilation).
As seen in the figure,
$\eta_{\nu}$ in the Newtonian potential is much larger than
that in PW potential. For $j=0$, the former can reach
a maximum value of $0.206$ at $\dm = 0.45$, which is far beyond
the maximum possible efficiency in the Schwarzschild geometry
($\eta = 0.057$) and is unphysical.
In fact, by integrating the viscous heating $\Qvis$ from $3 \Rg$
to the infinite outer boundary of the flow, we can obtain the
theoretical maximum $\eta_\nu$ for the above three solutions:
$1/4$ for the Newtonian potential with $j=0$ [from Eq.~(14)
of DPN], $1/12$ for the Newtonian potential with $j=1.225 c\Rg$
[from Eq.~(32) of Kohri \& Mineshige 2002],
and $1/16$ for PW potential with $j=1.837 c\Rg$.
Obviously the result with PW potential is the closest to
the reality ($0.057$), while the results with the Newtonian potential
are unreasonable.

We conclude for the moment that the usage of the Newtonian potential
is invalid in calculations for NDAFs at least at the following two points:
(1) it would overestimate substantially the optical depth for neutrinos;
(2) it would lead to an unphysical efficiency of energy release by
neutrino radiation.

\section{Effect of the optical depth on the neutrino annihilation
luminosity}

Our method for calculating neutrino annihilation is similar to
many previous works (e.g., Ruffert et al. 1997; PWF;
Rosswog, Ramirez-Ruiz, \& Davies 2003).
Figure~2 shows the variations of $\Lv$ (the upper thin solid line) and
$\Lvv$ (the lower thick solid line) with $\dm$, where $\Lv$ is the luminosity
of neutrino radiation before annihilation, and
$\Lvv$ is the luminosity of neutrino annihilation
(which is the most important from the observational point of view),
both of them are calculated with PW potential.
The circles and triangles represent the results of PWF
for $\Lv$ (empty) and $\Lvv$ (filled), respectively.
It is seen that our results agree very well
with that of PWF for $\dm \la 1$, because PW potential is a
good approximation for the Schwarzschild geometry.
For $\dm > 1$, our results
are lower than that of PWF. This is because they assumed neutrinos
to be optically thin; while we use the bridging formula for $\Qv$,
and there exists an optically thick region for $\dm > 1.2$.
According to our calculations
$\Lvv$ varies from $3.9 \times 10^{50}$ \ergs to
$3.6 \times 10^{52}$ \ergs for $1 < \dm < 10$,
which implies that, based on the energy consideration, NDAF can
indeed work as the central engine for GRBs.
In particular, our $\Lvv$ ($3.6 \times 10^{52}$ \ergs) for $\dm = 10$ 
is in good agreement with PWF's ``actual luminosity"
($\sim 4 \times 10^{52}$ \ergs, as mentioned in \S~1).

For comparison, Figure~2 also shows $\Lvv$ in other three cases:
(1) using PW potential but omitting the contribution from the optically
thick region (the dotted line, the $\tau > 2/3$ region appears for
$\dm > 1.2$);
(2) using the Newtonian potential and including the contribution from the
optically thick region (the dot-dashed line);
and (3) using the Newtonian potential but omitting that contribution
(the dashed line, the $\tau > 2/3$ region appears for $\dm > 0.052$).
As known from \S~3, the results of cases (2) and (3) are unreal because
the usage of the Newtonian potential overestimates unphysically
both $\tau$ and $\Lv$. It is seen that the omitting of the
contribution from the $\tau > 2/3$ region reduces substantially
$\Lvv$ in case (3), i.e., even with the overestimated $\Lv$ caused
by the Newtonian potential; as well as in case (1), i.e., even
the general relativistic effect is considered.
This is probably the reason why DPN obtained $\Lvv$ in their
Newtonian calculations which is insufficient for GRBs.
We think that it is unfair to ignore totally the neutrino radiation
from the $\tau > 2/3$ region. As DPN also stated, the neutrino emission
is partially suppressed as the inner regions of the flow are
becoming opaque. The trapping of neutrinos is a process that
is strengthening gradually with increasing $\tau$, the value of
$\tau$ reaching 2/3 does not mean that all neutrinos are suddenly
trapped, and the use of the bridging formula for $\Qv$ is exactly
to calculate the neutrino radiation from both the optically thin
and thick regions. A similar bridging formula has been widely used
for calculating the radiation of photons in both the optically thin
and thick cases (e.g., Narayan \& Yi 1995).

\section{Energy advection}

DPN argued that energy advection would become the dominant
cooling mechanism when the flow is optically thick for neutrinos.
As seen from their Fig.~3, however, it is not the case.
For example, for $\dm = 1$ the flow is optically thick at $R = 10\Rg$,
but the advection factor $f_{\rm adv} \equiv \Qadv/\Qvis$ at
this radius is only $\sim 0.1$.
In our opinion, whether cooling is dominated by advection or by radiation
is not determined by the optical depth.
For accretion flows in X-ray binaries and AGNs,
it is known that $f_{\rm adv}$ is relevant to the geometrical depth
rather than the optical depth of the flow
(Abramowicz, Lasota, \& Xu 1986):
\beq
f_{\rm adv} \propto (\frac{H}{R})^2 \ .
\eeq
Such a relationship can be well checked by the four
representative types of accretion models:
the optically thick standard thin disk (Shakura \& Sunyaev 1973)
and the optically thin SLE disk (Shapiro, Lightman, \& Eardley 1976)
are both geometrically thin and radiation-dominated,
i.e., $f_{\rm adv} \sim 0$;
and the optically thick slim disk (Abramowicz et al. 1988) and 
the optically thin advection-dominated accretion flow
(Narayan \& Yi 1994) are both geometrically thick and
advection-dominated, i.e., $f_{\rm adv} \sim 1$.
We argue that this relationship should also work
in the NDAF model with the following modification
in accordance with PW potential:
\beq
f_{\rm adv} \propto f_H \equiv (\frac{H}{R})^2 f^{-1} g^{-1} \ ,
\eeq
where $f_H$ is called by us the geometrical depth factor,
and $f^{-1} g^{-1}$ comes from the expression of $\Qvis$ [Eq.~(5)].

As shown in Figure~3, the variation of $f_{\rm adv}$ with $R$
(the solid line) agrees very well with that of $f_H$
(the dot-dashed line), but differs significantly from
that of $\tau$ (the dotted line),
clearly indicating that the strength of energy advection is relevant
to the geometrical depth rather than the optical depth.
It is also seen that, although
the flow is optically thick for $R < 25.9 \Rg$,
advection dominates over neutrino cooling
($f_{\rm adv} > f_{\nu}$, where $f_{\nu} \equiv \Qv /\Qvis$
is drawn by the dashed line) only in a smaller region $R < 5.1 \Rg$.
Once again, this result supports the view that it is important
to consider the role of the optically thick,
but neutrino radiation-dominated region,
e.g., $5.1 \Rg < R < 25.9 \Rg$ in the example of Figure~3.

\section{Discussion}

We have shown that the usage of the Newtonian potential along with
the omitting of neutrino radiation from the optically thick region
would lead to unreal luminosities for NDAFs, and that when the general
relativistic effect is considered and the contribution from the
optically thick region is included, NDAFs can work as the central
engine for GRBs.

In addition to its mass, a black hole may have its spin as the
other fundamental property. We consider here only the nonrotating
black hole, for which PW potential can work. PWF has shown that
a spining (Kerr) black hole will enhance the efficiency of
neutrino radiation. This strengthens our conclusion here that
NDAFs into black holes can be the central engine for GRBs.

We have tried to update partly the NDAF model by considering both the
effects of general relativity and neutrino opacity. There are certainly
other factors which may influence the neutrino luminosity of an NDAF
and we do not consider here, such as the electron degeneracy. We adopt
a simple treatment for the electron degeneracy pressure in agreement
with PWF and DPN. Kohri \& Mineshige (2002) pointed out that it is
important to include the effect of electron degeneracy that suppresses
the neutrino cooling at high density and high temperature.
Most recently, Kohri et al. (2005) considered the effects of both
electron degeneracy and neutrino optical depth, and calculated the
neutrino cooling, the electron pressure, and other physical quantities
even in the delicate regime where the electron degeneracy is moderate;
while in previous works as well as ours here the calculations can be
made accurate only in the two opposite limits of extremely degenerate
electrons and fully nondegenerate electrons. We wish to see in future
studies how the electron degeneracy would affect our results here.

\acknowledgments

We thank Ye-Fei Yuan for beneficial discussion and the referee
for helpful comments.
This work was supported by the National Science Foundation of China
under Grant Nos.10233030, 10503003 and the Natural Science Foundation
of Fujian Province under Grant No.Z0514001.

\clearpage

\begin{figure}
\plottwo{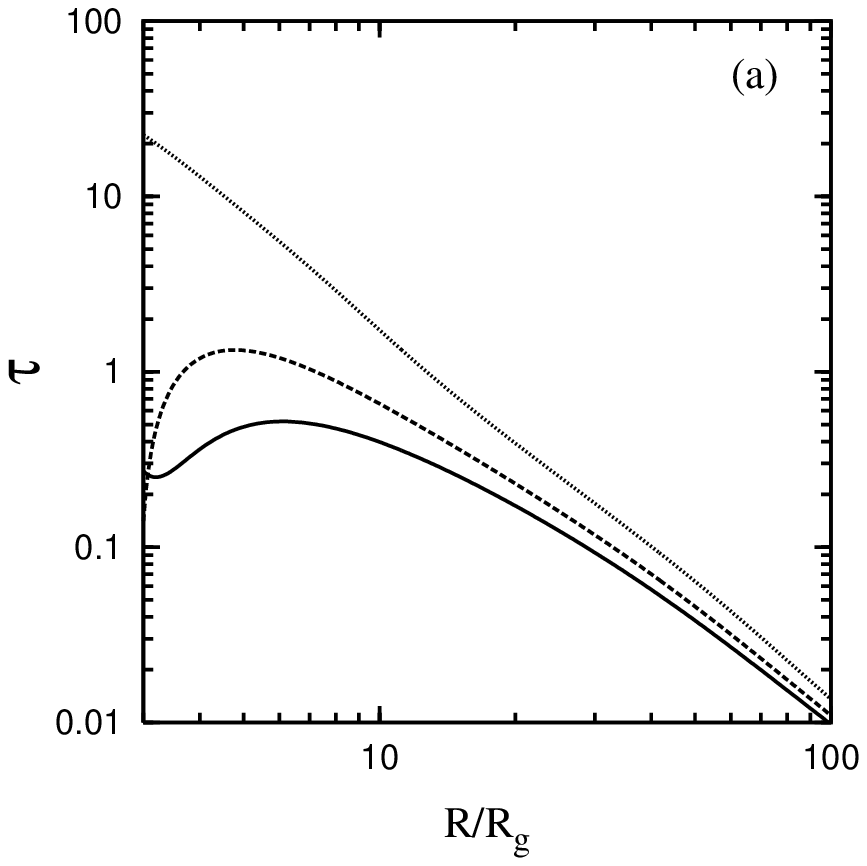}{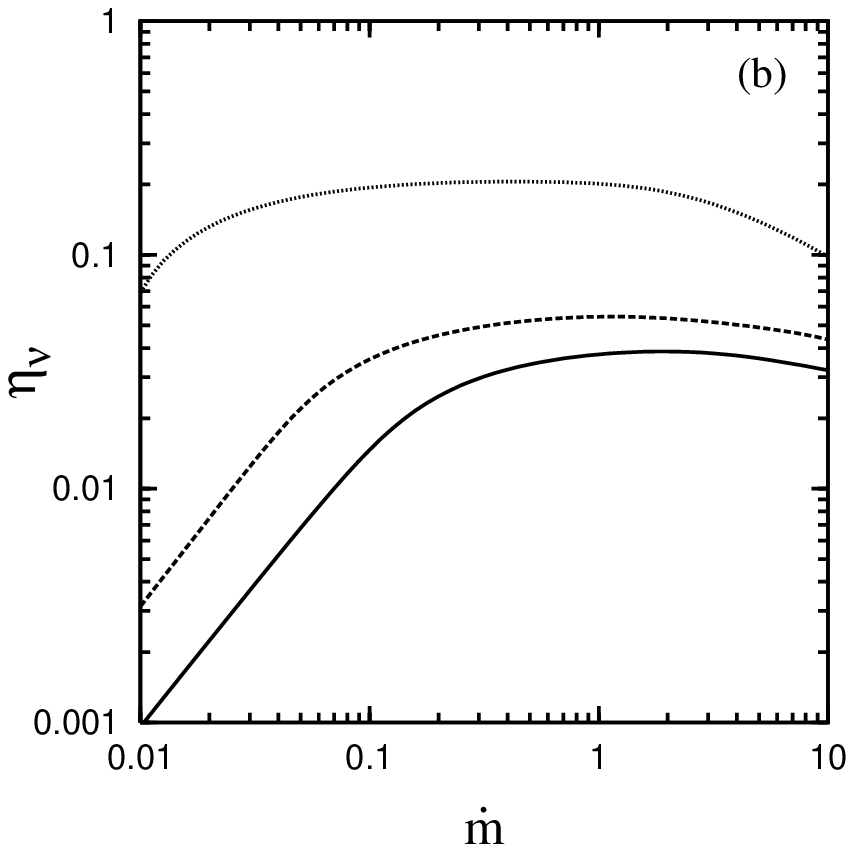}
\caption{
Solutions in PW potential with the accreted specific angular momentum
$j = 1.8 c\Rg$ (the solid line), in the Newtonian potential with
$j = 1.2 c\Rg$ (the dashed line), and in the Newtonian potential with
$j = 0$ (the dotted line). (a) Neutrino optical depth $\tau$
as a function of radius $R$ for the dimensionless mass accretion rate
$\dm = 1$; (b) Efficiency of neutrino radiation $\eta_{\nu}$
as a function of $\dm$.
\label{fig1}}
\end{figure}

\clearpage

\begin{figure}
\plotone{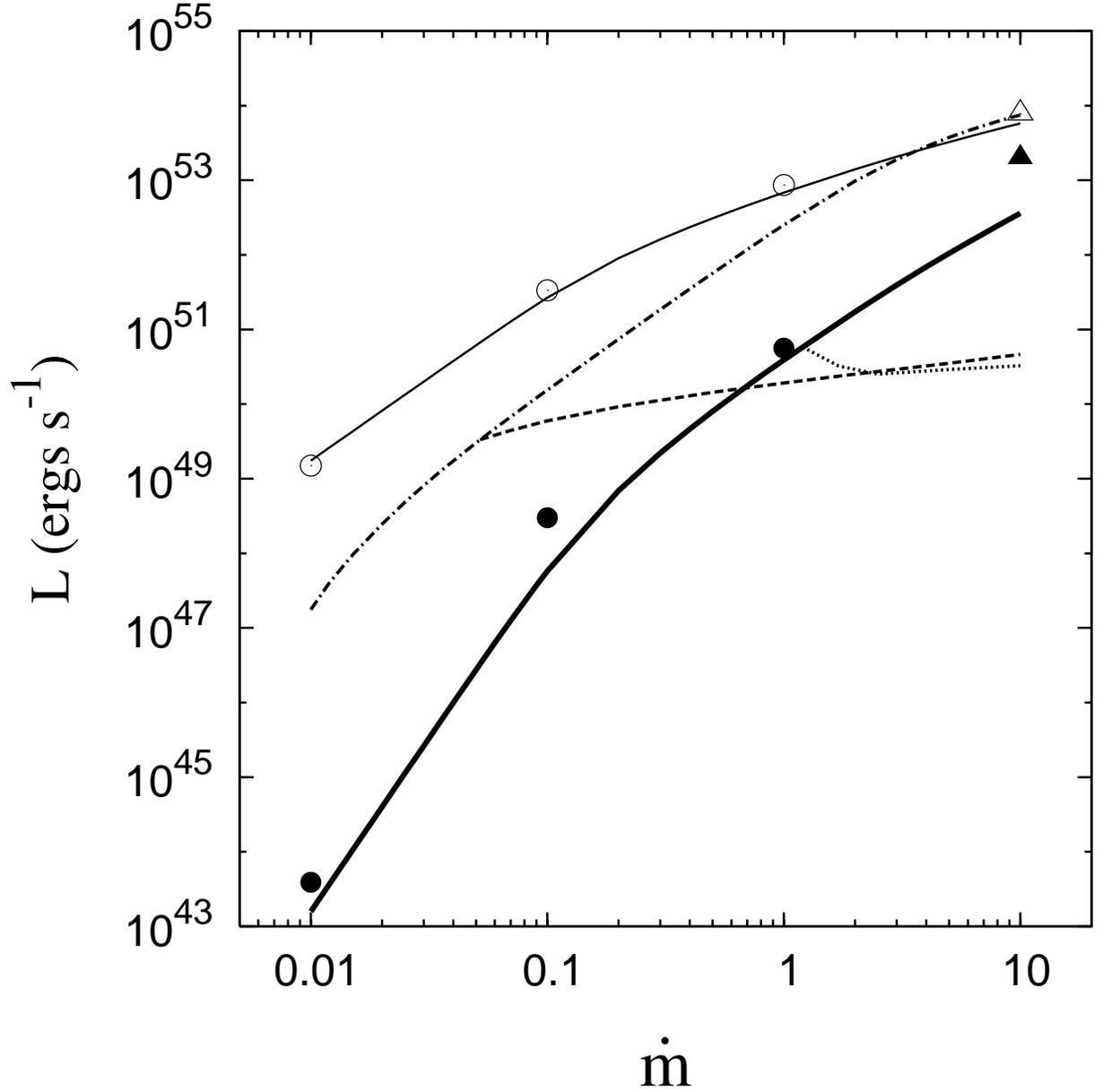}
\caption{
Neutrino luminosity before annihilation $\Lv$ with PW potential
(the thin solid line), neutrino annihilation luminosity $\Lvv$
with PW potential and including the $\tau > 2/3$ region
(the thick solid line), $\Lvv$ with PW potential but omitting
the $\tau > 2/3$ region (the dotted line), $\Lvv$ with the Newtonian
potential and including the $\tau > 2/3$ region (the dot-dashed line),
and $\Lvv$ with the Newtonian potential but omitting the $\tau > 2/3$
region (the dashed line) as functions of $\dm$.
The empty circles and triangle denote $\Lv$ of PWF,
and the filled circles and triangle denote $\Lvv$ of PWF.
\label{fig2}}
\end{figure}

\clearpage

\begin{figure}
\plotone{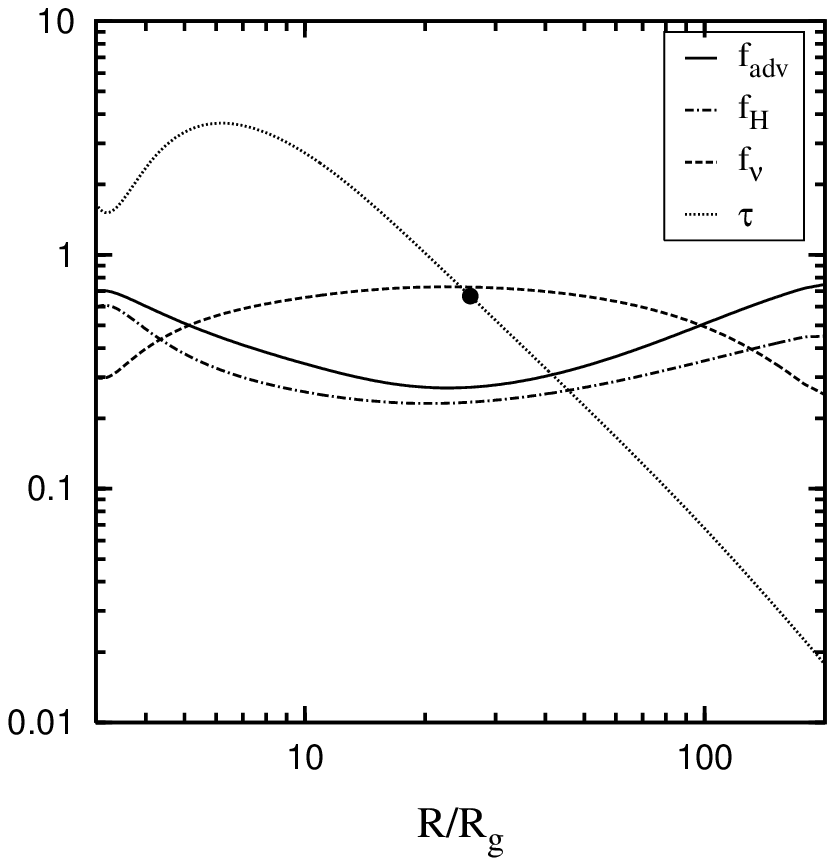}
\caption{
Advection factor $f_{\rm adv}$, geometrical depth factor $f_H$,
neutrino cooling factor $f_{\nu}$, and $\tau$ as functions
of $R$ for $\dm = 5$. The filled circle denotes the $\tau = 2/3$
position, i.e., $R = 25.9 \Rg$.
\label{fig3}}
\end{figure}


\begin{thebibliography}{}
\bibitem[]{} Abramowicz, M. A., Czerny, B., Lasota, J.-P., \&
Szuszkiewicz, E. 1988, \apj, 332, 646
\bibitem[]{} Abramowicz, M. A., Lasota, J.-P., \& Xu, C. 1986, in
IAU Symp. 119 Quasars, ed. G. Swarup and V. K. Kapahi
(Dordrecht: Reidel), 376
\bibitem[]{} Di Matteo, T., Perna, R., \& Narayan, R. 2002, \apj,
579, 706 (DPN)
\bibitem[]{} Kohri K., \& Mineshige S. 2002, \apj, 577, 311
\bibitem[]{} Kohri K., Narayan, R., \& Piran, T. 2005, \apj, 629, 341
\bibitem[]{} M\'esz\'aros, P. 2002, \araa, 40, 137
\bibitem[]{} Narayan, R., Paczy\'nski B., \& Piran, T. 1992, \apj, 395, L83
\bibitem[]{} Narayan, R., Piran, T., \& Kumar, P. 2001, \apj, 557, 949
\bibitem[]{} Narayan, R., \& Yi, I. 1994, \apj, 428, L13
\bibitem[]{} Narayan, R., \& Yi, I. 1995, \apj, 452, 710
\bibitem[]{} Paczy\'nski, B. \& Wiita, P. J. 1980, \aap, 88, 23
\bibitem[]{} Popham, R., Woosley, S.E., \& Fryer, C. 1999, \apj,
518, 356 (PWF)
\bibitem[]{} Rosswog, S., Ramirez-Ruiz, E., \& Davies, M. 2003, \mnras,
345, 1077
\bibitem[]{} Ruffert, M., Janka, H.-T., Takahashi, K., Schaefer, G.
1997, \aap, 319, 122
\bibitem[]{} Shakura, N. I., \& Sunyaev, R. A. 1973, \aap, 24, 337
\bibitem[]{} Shapiro, S. L., Lightman, A. P., \& Eardley, D. N. 1976,
\apj, 204, 187
\bibitem[]{} Zhang, B., \& M\'esz\'aros, P. 2004, Int. J.
Mod. Phys. A, 19, 2385
\end{thebibliography}
\end{document}